%% file: main.tex
\documentclass[12pt]{article}
\RequirePackage{amsthm,amsmath,amsfonts,amssymb}
\RequirePackage[authoryear]{natbib}
\RequirePackage[colorlinks,citecolor=blue,urlcolor=blue]{hyperref}

\RequirePackage{algorithm, algorithmic}
\RequirePackage{graphicx}
\RequirePackage{multirow}
\RequirePackage{threeparttable}
\usepackage[margin=1in]{geometry}

\input{mydef.tex}

\begin{document}

\title{Bayesian Kernel Machine Regression via Random Fourier Features for Estimating Joint Health Effects of Multiple Exposures}

\author{Danlu Zhang$^{1,*}$, Stephanie M. Eick$^{2}$, and 
Howard H. Chang$^{1,2}$\\
$^{1}$Department of Biostatistics and Bioinformatics, Emory University, USA \\
$^{2}$Gangarosa Department of Environmental Health, Emory University, USA
\\
$^*$danlu.zhang@emory.edu}

\date{}

\maketitle

\centerline{\textbf{Abstract}}
\vtwo

Environmental epidemiology has traditionally examined single exposure one at a time. Advances in exposure assessment and statistical methods now enable studies of multiple exposures and their combined health impacts. Bayesian Kernel Machine Regression (BKMR) is a widely used approach to flexibly estimates joint, nonlinear effects of multiple exposures. But BMKR is computationally intensive for large datasets, as repeated kernel inversion in Markov chain Monte Carlo (MCMC) can be time-consuming and often infeasible in practice. To address this issue, we propose using supervised random Fourier basis functions to replace the Gaussian process random effects. This re-frames the kernel machine regression into a linear mixed-effect model that facilitates computationally efficient estimation and prediction. Bayesian inference is conducted using MCMC with Hamiltonian Monte Carlo algorithms. Simulation studies demonstrate that our method yields results comparable to BKMR while significantly reduces the computation time. Our approach outperforms BKMR when the exposure-response surface has stronger dependency and when using predictive process as an alternative approximation method. Finally, we applied this approach to analyze over 270,000 birth records, examining associations between multiple ambient air pollutants and birthweight in Georgia.

\vtwo



\section{Introduction}
\label{s:intro}

Most environmental epidemiological studies focus on estimating the health effects of a single exposure one-at-a-time, sometimes adjusting for other exposures. However, humans are exposed to multiple environmental risk factors simultaneously and the majority of exposures are correlated due to common sources or exposure pathways (\cite{BILLIONNET2012126}). Recent and continual developments of statistical approaches have facilitated our ability to examine joint effects of multiple exposures (\cite{BENKACOKER2020109903,sexton2012cumulative}). For example, previous studies have documented joint effects of multiple air pollutants on adverse health outcomes, such as bladder cancer incidence (\cite{kim2024association}), childhood persistent asthma (\cite{shiroshita2024joint}) and low birthweight (\cite{yang2020associations}). 

Bayesian kernel machine regression (BKMR) is one widely used method to estimate joint effects of multiple exposures (\cite{10.1093/biostatistics/kxu058}). Compared to other commonly used models, such as weighted quantile sum regression (WQS) and quantile g-computation (q-gcomp), BKMR assumes that the complex nonlinear exposure-response surface is a random realization of a Gaussian process specified by its kernel function (\cite{10.1093/biostatistics/kxu058}). Hence, one can easily derive marginal effect of any exposure combination, offering a flexible way to evaluate health effects without pre-specifying exposure contrasts of interest (\cite{10.1093/biostatistics/kxu058,li2022health}). BKMR has also been shown to provide reliable inference in complex and high-dimensional exposure settings. 

Kernel regression models are known to be computationally intensive due to the need to invert the $n \times n$ kernel matrix, where $n$ is the sample size. This issue is exacerbated in BKMR due to its use of Markov Chain Monte Carlo (MCMC) for posterior inference. Currently, BKMR can be impractical for moderately large datasets (e.g. with sample sizes around 5,000) and as a result, BKMR is more commonly used in analyzing health effects of chemical exposures from cohort studies with a relatively small sample size (\cite{eick2024prenatal,preston2020prenatal,zhuang2021effects}). There is a pressing need to address this computational challenge for analysis of large administrative health databases where sample sizes are often in the tens of thousands or larger. Current implementation of BKMR offers an option to use predictive process where exposures are projected to a lower dimensional space spanned by the set of knots (\cite{10.1111/j.1467-9868.2008.00663.x,bobb2018statistical}). However, this approach has been noted for challenges with overs-smoothing (\cite{10.1214/17-BA1056R,10.1016/j.csda.2009.02.011}), and there has been little guidance on knot selection in multivariate Gaussian process settings. One possibility is to use a grid of exposure quantiles as knots (e.g., all combinations of exposure deciles), but the number of knots increases quickly as the number of exposures included in the exposure matrix increases. 

One common way to reduce the memory requirement and computational cost is with low-rank approximation, which is widely used in large-scale data applications (\cite{kishore2017literature}). According to the spectral representation theorem, trigonometric functions can be used as basis functions to capture essential characteristics of the data with fewer parameters (\cite{MILLER2022100598}). Random Fourier features (RFF) with frequencies sampled from a known spectral density can be used to approximate Gaussian process in a kernel machine setting (\cite{rahimi2007random}). This allows for the application of kernel methods with linear algorithms, significantly improving scalability while maintaining quality. Although the conventional implementation of RFF is a data-independent approach, it has been extended to be data-driven (supervised) in several ways for large real-world applications (\cite{chang2017data,liu2021random}). RFF has also recently been examined for analyzing large spatial dataset (\cite{MILLER2022100598}).

The main objective of this paper is to adopt and evaluate the use of RFF for Gaussian process random effects in BKMR to reduce the computation time. We chose this low-rank method because this re-frames the kernel machine regression into a linear mixed-effect model that facilitates computationally efficient estimation and prediction under a Bayesian framework. Compared to other basis functions used in spatial statistics, such as the compact Wendland (\cite{nychka2015}), RFF more readily extends to higher dimensions. Also, in a supervised framework, RFF allows the basis functions to be adaptive. This is in contrast the Karhunen–Loève expansion, which is deterministic given the kernel parameters. The rest of the paper is organized as follows. In Section 2, we introduce the proposed method, which we call Fast BKMR. In Section 3, we present simulation studies and compare our results with BKMR when possible under different settings. In Section 4, we apply Fast BKMR to characterize associations between birthweight and multiple ambient air pollutions to highlight the ability of Fast BKMR to handle a large sample size and to capture complex exposure-response functions. In Section 5, we discuss the advantages and limitations of this method and potential further extensions.

\section{Methods}
\label{s:method}

\subsection{Bayesian Kernel Machine Regression for Multiple Exposures}

We first describe the current formulation of BKMR for multiple exposures (\cite{10.1093/biostatistics/kxu058}). For individual $i=1, \ldots n$, let $Y_i$ denote the outcome of interest and $\bx_i = (x_{1,i}, \ldots, x_{M,i})$ denote the vector of $M$ exposures. We assume the following linear regression model
\begin{equation}
Y_i = h(\bx_i)+\bz^T_i\bgamma + \epsilon_i
\end{equation}
where $h(\bx_i)$ is the individual-specific association between $\bx_i$ and the outcome, $\bz_i$ is the vector of confounders with corresponding regression coefficient $\bgamma$, and $\epsilon_i \stackrel{\text{i.i.d.}}{\sim} N(0, \sigma^2)$ is the residual error. BKMR represents $h(\bx)$ non-parametrically as a realization from a Gaussian process, characterized by the kernel
\begin{equation}
     \mathcal{K} (\, \bx_i, \bx_j\,) = \exp\left( -\sum_{m=1}^M\theta_m  (x_{m,i}-x_{m,j})^2\right)
\end{equation}

Hence, effects of multiple exposures are more similar for individuals with smaller squared Euclidean distance between exposure levels. The above kernel also assumes separability between exposures, such that the correlation decreases multiplicatively by exposures and the importance of each exposure is allowed to vary via the exposure-specific parameter $\theta_m \ge 0$.

The joint distribution of  $\bh = [ \, h(\bx_1), \ldots, h(\bx_n) \,]^T $ is multivariate normal with  covariance matrix $\tau^2\bK$, where $\tau^2$ is the marginal variance and $\bK$ is an $n\times n$ matrix with entries given by the kernel function. Bayesian inference via MCMC can be computationally challenging for large sample size $n$ due to the need to repeatedly evaluate $\bK^{-1}$.  

\subsection{Kernel Approximation by Random Fourier Features}

Bochner's theorem (\cite{bochner1955harmonic, rudin2017fourier}) states that any positive definite and stationary kernel can be expressed as the Fourier transform of its spectral density $f(\bomega)$:
\begin{equation}
    \mathcal{K} (\, \bx_i-\bx_j\,)  = \int \exp \{\,i \bomega^T(\bx_i - \bx_j) \}\,f(\bomega) \, d\bomega
\end{equation}
where $\bomega \in \mathbb{R}^M$ is the frequency and $i=\sqrt{-1}$. Consequently, the random effect $h(\bx_i)$ can be rewritten as
\begin{equation}
    h(\bx_i) = \int \exp \left(\,i \bomega^T\bx_i \right) \, \left[ a(\bomega) + ib(\bomega) \right] \, d\bomega
\end{equation}
where $a(\bomega)$ and $b(\bomega)$ are mean-zero random variables that are independent between $\bomega$ and have $\text{Var} \left[ a(\bomega) + ib(\bomega) \right] = \tau^2 f(\bomega)$ (\cite{gelfand2010handbook, lindgren2012stationary}). Since $h(\bx_i)$ is real-valued, the integral can be approximated by finite set of frequencies $\{ \bomega_1, \ldots, \bomega_J \}$, 
\begin{equation}
\label{h_basis}
    h(\bx_i) = \sum_{j=1}^J a_j \cos (\bomega^T_j\bx_i) + b_j\,\sin (\bomega^T_j\bx_i)
\end{equation}

Equation (\ref{h_basis}) can be viewed as a set of basis expansions using periodic functions with amplitudes $a_j$ and $b_j$ as random-effect coefficients to be estimated from the data. This RFF has been developed and applied on large spatial data in previous study (\cite{MILLER2022100598}). For the Gaussian process and kernel function specified by BKMR, these correspond to
$\bomega_j \sim N(\,\bZero, \Sigma\,)$ and $a_j, b_j \sim N (\, 0, \tau^2/J)$, where $\Sigma$ is an $m \times m$ diagonal matrix with elements $\{ 2\theta_1, \ldots, 2\theta_M\}$. This re-frames the kernel machine regression into a linear mixed-effect model that facilitates computationally efficient parameter estimation and prediction. It can be shown that marginalizing over $a_j$ and $b_j$, the induced $\text{Cov} \{ h(\bx_i), h(\bx_j) \}$ has expectation $\mathcal{K} (\, \bx_i, \bx_j\,)$ and the variance decreases as a function of $J$.  Moreover, by treating the frequency $\bomega_j$ as a random variable, we can learn the optimal finite number of frequencies to use from the data.

We complete the Bayesian hierarchical model specification by assigning improper flat prior for the regression coefficients $\bgamma$ with $N(0, \sigma_{\gamma}^2 \boldsymbol{I})$, inverse-Gamma distributions for the residual variance $\sigma^2\sim \text{IG} \,(0.001, 0.001)$, the Gaussian process marginal variance $\tau^2 \sim \text{IG}\, (0.001, 0.001)$, and $\theta_1, \ldots, \theta_M \sim \text{IG}\, (0.001, 0.001) $.

\subsection{Bayesian Inference and Estimation}

Hamiltonian Monte Carlo (HMC; \cite{neal2011handbook}) is an approach to generate posterior samples in high-dimensional space which is suitable to deal with the label-switching problem in Bayesian analysis (\cite{MILLER2022100598}). We implement HMC using a leapfrog integrator. Let $\mathcal{L}, e, \text{and}\ L$ denote the log-density, step-size parameter and number of steps, respectively. Starting at the initial location $\theta^0$, the idea of HMC at $m^{th}$ iteration is as follows. 

\begin{enumerate}
\item Sample $r^0$ from $N(0,I)$ and set $\theta^m \leftarrow \theta^{m-1}$, $\Tilde{\theta} \leftarrow \theta^{m-1}$ and $\Tilde{r} \leftarrow r^0$.
\item for $l = 1, \ldots, L$, set $\Tilde{\theta}, \Tilde{r} \leftarrow \text{Leapfrog}(\Tilde{\theta}, \Tilde{r},e)$.
\item Update $\theta^m$ and $r^m$ by $\Tilde{\theta}$ and $-\Tilde{r}$ with probability $\alpha = \text{min} \{1, \frac{\exp (\mathcal{L} (\Tilde{\theta}) - \frac{1}{2} \Tilde{r} \cdot \Tilde{r})}{\exp (\mathcal{L} (\theta^{m-1}) - \frac{1}{2} r^0 \cdot r^0)}\}$. 
\end{enumerate}

To tune $e$ to achieve the acceptance rate of $65-85\%$, we checked the acceptance rate per 200 iterations in burn-in samples. If the acceptance rate is over $85\%$, $e$ will be replaced by $e(1+e_t)$, where $e_t \in [0, 1]$ is the tuning rate. In contrast, if the acceptance rate is lower than $65\%$, $e$ will be set to $e(1-e_t)$. The $\text{Leapfrog}(\theta, r, e)$ is defined as:
\begin{enumerate}
\item Set $\Tilde{r} \leftarrow r + (e/2)  \nabla_{\theta} \mathcal{L}(\theta)$.
\item Set $\Tilde{\theta} \leftarrow \theta + e\Tilde{r}$.
\item Set $\Tilde{r} \leftarrow \Tilde{r} + (e/2)  \nabla_{\theta} \mathcal{L}(\Tilde{\theta})$.
\end{enumerate}

Bayesian inference is conducted by posterior sampling using Markov chain Monte Carlo (MCMC). We first initialize $\theta_1, \ldots, \theta_M $ and then simulate $\bomega_1, \ldots, \bomega_J $ from $N(0, \Sigma)$. Basis functions are constructed via $\cos(\bomega_j^T x_i)$ and $\sin(\bomega_j^T x_i), j = 1, \ldots, J, i = 1, \ldots, n$. Next, we utilize penalized regression to estimate $a_j, b_j$ and $\bgamma$. Last step in initialization is to estimate $\tau^2$ and $\sigma^2$. 
In MCMC iteration, Gibbs updates are used for $\sigma^2$, $\tau^2$ and $\theta_1, \ldots, \theta_M$. $(\bgamma, a_1, \ldots, a_J, b_1, \ldots, b_J)$ are updated via HMC as one block and $\bomega_1, \ldots, \bomega_J $ are also updated via HMC as another block. Log-likelihoods and first derivatives of log-likelihood used in the HMC updates can be found in Supplementary Material.

\begin{algorithm}
\caption{MCMC iterations of Fast BKMR}
\begin{algorithmic}
\STATE \textbf{Input:} Data: $\{Y_i, \bx_i, \bz_i\}_{i=1}^n,\,\ e_{\beta},\,\ e_{\omega},\,\ J,\,\ \sigma_{\gamma}^2,\,\ (\theta_1, \ldots, \theta_M),\,\ \tau^2,\,\ \sigma^2$.
\STATE \textbf{Initialization:}
\STATE Simulate $\bomega_j^0 \sim N(\bZero, \Sigma), j = 1, \ldots, J$ where $\Sigma = \text{diag} (2\theta_1, \ldots, 2\theta_M)$.
\STATE Construct basis functions $\cos(\bomega_j^{0T} x_i)$ and $\sin(\bomega_j^{0T} x_i)$.
\STATE Estimate $a_j^0,\,\ b_j^0$ and $\bgamma^0$ via penalized regression using $\sigma_{\gamma}^2,\,\ \tau^2,\,\ \sigma^2$ and input data.
\STATE Estimate $\sigma_0^2$.
\FOR {$k$ in $1, \dots, K$}
    \item Update $\theta_1^k, \ldots, \theta_M^k$ via Inverse Gamma $(0.001 + \frac{J}{2}, 0.001 + \frac{1}{2} \sum_{j=1}^J \bomega_j^{(k-1)T} \bomega_j^{(k-1)})$.
    \item Update $(\bgamma^k, a_1^k, \ldots, a_J^k, b_1^k, \ldots, b_J^k)$ as a block via HMC.
    \item Update $\bomega_1^k, \ldots, \bomega_J^k$ as a block via HMC.
    \item Update $\tau^2_k/J$ via Inverse Gamma $(0.001 + J, 0.001 + \frac{1}{2} \sum_{j=1}^J (a_j^k)^2 + (b_j^k)^2)$.
    \item Update $\sigma_k^2$ via Inverse Gamma $(0.001 + \frac{n}{2}, 0.001 + \frac{1}{2} \sum_{i=1}^n (Y_i - \hat{Y}_i^k)^2)$.
    \item Calculate $h(\bx_i^k)$ via $\sum_{j=1}^J a_j^k \cos (\bomega^{kT}_j\bx_i) + b_j^k \sin (\bomega^{kT}_j\bx_i)$.
\ENDFOR
\end{algorithmic}
\end{algorithm}

Watanabe-Akaike Information Criterion (WAIC) can be used to select the number of basis functions $J$. WAIC is helpful for models with hierarchical structures. WAIC is defined as $-2(\text{lppd} - p_{\text{WAIC}})$, where $\text{lppd}$ is the sum of log point-wise predictive densities and $p_{\text{WAIC}}$ denotes the sum of posterior variance of the log predictive density for each data point $Y_i$ (\cite{gelman2013bayesian}). We discarded the first half of Markov chain Monte Carlo samples, and inference was conducted using the second half of posterior samples. All computations were conducted in R version 4.2.2 (\cite{R}) and the R package, bkmr version 0.2.2, was used to fit BKMR (\cite{bkmr2022}). We also provide example R code for implementing Fast BKMR in Supplementary Material.

\section{Simulation Studies}
\label{s:simulation}

\subsection{Setup}

We generated datasets of $N$ observations $\{Y_i, \bx_i, \bz_i\}_{i=1}^N$, where $\bx_i = (x_{i1}, \ldots, x_{iM})^T$ represents an exposure profile with $M$ exposures and $\bz_i = (z_{i1}, \ldots, z_{iP})^T$ denotes $P$ confounders for individual $i$. We examined 2, 5 and 10 exposures and all exposures were simulated from normal distributions with mean zeros and different variances. Five confounders were generated from either the normal distribution or binomial distributions. Details of the simulated exposures and confounders are provided in Web Table 1. $h(\bx_i)$ was simulated from a Gaussian process with kernel $\bK$ with different levels of correlation: strong (correlations between 0.75 and 0.9) and weak (correlations between 0.1 and 0.3). Outcome $Y_i$ was simulated from $N(h(\bx_i) + \bz_i^T \bgamma, \sigma^2)$. The weak correlation scenario is designed to be a particularly challenging setup where the exposure-response surface is highly variable. We also examined scenarios where BKMR incorrectly specifies the kernel, with the true kernel being either $\exp (- \sum_{m=1} ^M \theta_m \sqrt{|x_{im} - x_{jm}|})$ or $\exp (- \sum_{m=1} ^M \theta_m |x_{im} - x_{jm}|)$. We compared our model to BKMR and BKMR with predictive process with sample sizes $N$ of 200, 500 and 1000 across 100 simulations. To further evaluate computation time, we applied BKMR with predictive process and Fast BKMR to datasets with 5,000 and 10,000 observations. Because the computation time was consistent for each simulation, we ran 10 simulations for these large sample sizes. 

Within each simulated dataset, different number of basis functions ($J$ = 5, 20, 100, 200, 500, and 1000). Considering prior authors used 50 and 100 knots for 7 exposures in the BKMR software article (\cite{bobb2018statistical}), the number of knots in predictive process was set as ten times of exposures in our simulation studies, which were 20, 50 and 100 for 2, 5 and 10 exposures. We used the estimated $\theta_m, m = 1, \ldots, M$ obtained from BKMR R package using a sub-sample of 100 observations as the initial values for $\theta_m$ in the Fast BKMR method. Model performance was evaluated by comparing the root mean squared error (RMSE) between the estimated $\hat{h}(\bx_i)$ and the true $h(\bx_i)$.

We also evaluated the model's ability to perform out-of-sample prediction. We generated 20\%, 30\% and 50\% more data and left those out as testing data. The RMSE between the interpolated exposure-response function $\Tilde{h}(\bx_i^{new})$ and the true $h(\bx_i)$ was calculated to evaluate model's ability.

\subsection{Results}

\begin{figure}
\centering
\includegraphics{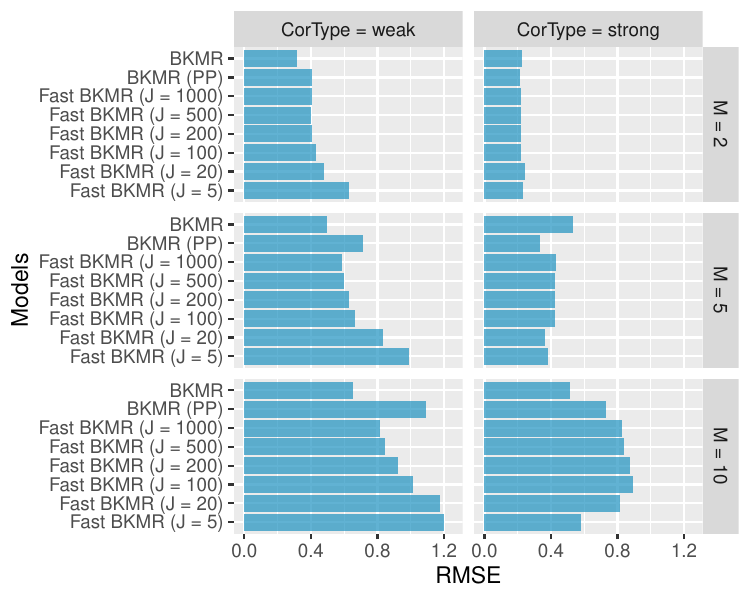}
\caption{The RMSE between true and estimated joint effects with different correlation type (CorType) and number of exposures ($M$) at sample size of 1000.}
\label{figure:H hat rmse}
\end{figure}

Simulation results of RMSE for in-sample tests are presented in Figure~\ref{figure:H hat rmse} for sample size of 1,000. When the correlation between $h(\bx_i)$ is weak, using RFF results in higher RMSE, which can be improved by increasing the number of basis functions. This is expected because of the highly variable exposure-response function and BKMR is the true data generating process. However, as an approximation for Gaussian process, fast BKMR outperforms predictive process even with a small number of basis functions. When the correlation between $h(\bx_i)$ is strong, Fast BKMR outperformed BKMR when using a smaller number of basis functions. This is likely because optimal frequencies can be learned from the data to effectively model less variable exposure-response functions. As the number of exposures increases, RMSE also increases. We observed similar patterns between methods for out-of-sample evaluation and other sample sizes (Web Figure 1 and Figure 2 A).  

In the presence of kernel misspecifications, all methods resulted in larger RMSE values across all combinations of sample size, number of exposures and correlation types. Smaller RMSE was associated with stronger correlation between $h(\bx)i)$, smaller number of exposures and larger sample sizes. Again Fast BKMR performed better in scenarios with strong correlations between $h(\bx_i)$. For out-of-sample evaluation with incorrect kernel specifications, our method also had a smaller RMSE in scenarios with larger number of exposures (Web Figure 2 B).

Fast BKMR provided considerable reduction on computation time as shown in Table~\ref{table:compute_time}. The computation time of Fast BKMR included the time required to obtain initial values. Compared with BKMR using predictive process, our method reduced computation time by 84\% to 99\% when the number of knots and basis functions were the same. Our method was slightly slower when both the sample size and number of exposures were small (Web Table 2), but the reduction in computation time increased rapidly as sample size and the number of exposures grew. With a sample size of 10,000, our method required over 50\% less time for 5 or 10 exposures. With 10 exposures, BKMR with predictive process took 6.6 days to fit one model, whereas our method required around 1 day with 1,000 basis functions. We also observed that computation time of BKMR with predictive process was dependent to both sample size and number of exposures. In contrast, the computation time of our method increased almost linearly with sample size and the number of basis functions, while the number of exposures had only a minor effect. This indicates the increased applicability of Fast BKMR in settings with a large number of exposures.

\begin{table*}
\caption{Computation time in minutes (percentages of reduction$^\text{1}$) for BKMR, BKMR with Predictive Process (PP) and Fast BKMR.}
\label{table:compute_time}
\tabcolsep=0pt
\begin{tabular*}{\textwidth}{@{\extracolsep{\fill}}lllll@{}}
\hline
Sample Size & Type & $M^\text{2} = 2$ & $M^\text{2} = 5$ & $M^\text{2} = 10$ \\
\hline
\multirow{6}{5em}{1000} 
& BKMR & 833.12 (-5913\%)  &589.45 (-1386\%)  &902.41 (-582\%)  \\
& BKMR (PP) & 13.86 (0\%)  &39.67 (0\%)  &132.34 (0\%)   \\
& Fast BKMR (J = 5)   & 0.9 (94\%)  &1.23 (97\%)  &1.83 (99\%)  \\
& Fast BKMR (J = 100)  & 9.88 (29\%)  &12.24 (69\%)  &15.15 (89\%) \\
& Fast BKMR (J = 1000) & 106.05 (-665\%)  &126.45 (-219\%)  &151.96 (-15\%)  \\
\hline
\multirow{6}{5em}{5000$^\text{3}$} 
& BKMR  & --  & --  & --  \\
& BKMR (PP) & 326.8 (0\%)  &1039.44 (0\%)  &2393.1 (0\%)  \\
& Fast BKMR (J = 5)   & 3.53 (99\%)  &4.2 (100\%)  &3.88 (100\%)  \\
& Fast BKMR (J = 100)  & 59.13 (82\%)  &67.24 (94\%)  &60.14 (97\%) \\
& Fast BKMR (J = 1000) & 636.31 (-95\%)  &756.47 (27\%)  &766.56 (68\%)  \\
\hline
\multirow{6}{5em}{10000$^\text{3}$} 
& BKMR  & --  & --  & --  \\
& BKMR (PP) & 1428.39 (0\%)  &2945.9 (0\%)  &9478.92 (0\%)   \\
& Fast BKMR (J = 5)   & 6.76 (100\%)  &5.65 (100\%)  &7.32 (100\%) \\
& Fast BKMR (J = 100)  & 121.1 (92\%)  &100.79 (97\%)  &123.43 (99\%)  \\
& Fast BKMR (J = 1000) & 1382.06 (3\%)  &1294.75 (56\%)  &1473.09 (84\%)  \\
\hline
\end{tabular*}

\begin{tablenotes}
\item[$^{1}$] Percentage of reduction: $1 - \frac{\text{computation time of current model}}{\text{computation time of BKMR with PP}}$
\item[$^{2}$] $M$: the number of exposures. 
\item[$^{3}$] Average computation time was calculated across 10 simulations. 

\end{tablenotes}

\end{table*}

We examined the use of WAIC to select the number of basis functions and results are shown in Web Figure 3. Small number of basis functions were more often preferred for the kernel with strong correlation With increasing sample size and number of exposures, more basis functions were preferred. For incorrect specification of the kernel, less basis functions were preferred. 

\subsection{Parametric Exposure-Response Function}

In real-world applications, associations between exposures and outcome may not arise from a  Gaussian process. Hence, we also examined our method where a parametric form $h(\bx_i) = -10 + 2sin(x_{1,i}x_{2,i}) + 4 (x_{3,i} - 0.5)^2 + 2 x_{4,i} + x_{5,i}$ proposed by Friedman \cite{friedman1991} was used. We simulated $h(\bx_i)$ under two scenarios. In the first, 5 exposures contributed to $h(\bx_i)$; in the second, there were 10 exposures but only the first five contributed to $h(\bx_i)$, with the remaining exposures acting as noise. BKMR, BKMR with predictive process and our method were tested on datasets of 200, 500 and 1000 observations. For BKMR with predictive process, we used 50 and 100 knots for 5 and 10 exposures cases, respectively. For Fast BKMR, we employed 20, 200, 500 and 1000 basis functions.

\begin{figure}
\centering
\includegraphics{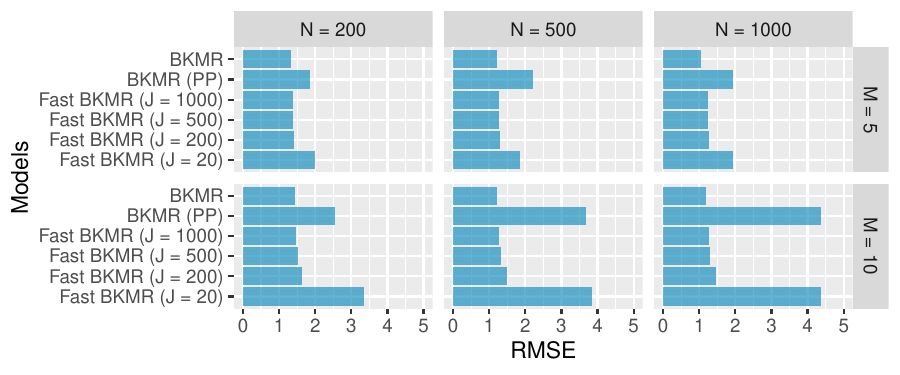}
\caption{The RMSE between true and estimated joint effects under parametric scenario with different sample size ($N$) and number of exposures ($M$).}
\label{figure:para rmse}
\end{figure}

Simulation results of RMSE for the parametric $h(\bx_i)$ are presented in Figure~\ref{figure:para rmse}. As the number of basis functions increased, Fast BKMR achieved lower RMSE, and with 1,000 basis functions, Fast BKMR performed comparable to BKMR. BKMR with predictive process performed similarly to Fast BKMR with 20 basis functions when there were 5 exposures. However, when 10 exposures were considered with only 5 of them contributing to $h(\bx_i)$, the performance of BKMR with predictive process was significantly affected by noise. In contrast, our method with basis functions over 200 demonstrated greater resilience to noise in the exposure data, indicating more stable estimation results.

\section{Application to the Atlanta Birthweight and Air Pollution Study}
\label{s:birth}
\subsection{Dataset}

We analyzed data from a previous study of associations between air pollution and birthweight (\cite{strickland2019associations}). The birth records from 20 counties in metropolitan Atlanta were obtained from the Office of Health Indicators for Planning, Georgia Department of Public Health for the period 2002 to 2006. A total of 273,711 singleton live birth with gestational weeks of 28 to 44 were included. We excluded birthweight less than 400 grams, maternal age below 16 or over 43 years of age, presence of any congenital anomaly identified on birth record and preterm birth (gestational weeks $\leq$ 36) with a procedure code for induction of labor. The characteristics of study population is shown in Web Table 3.

We considered three air pollutants: 1-hour maximum carbon monoxide (CO) and nitrogen dioxide (NO$_2$), and 24-hour average particular matter $\leq2.5$ $\mu$m in diameter (PM$_{2.5}$). Air pollutant concentrations were obtained by bias-correcting 12-km grid numerical simulations from the Community Multi-scale Air Quality (CMAQ) Model with ground monitor measurements in Georgia (\cite{doi:10.1021/acs.est.5b05134}). Birth records were linked to CMAQ grid cells using maternal address at delivery at the Census tract level, and average exposures were calculated for the entire pregnancy. A summary of exposure concentrations in original scale is shown in Web Table 4. Due to the varying ranges of the pollutants, we standardized pollutant concentrations by dividing their standard deviation. 

\subsection{Health Models}

To compare the results of Fast BKMR with other commonly used models, we also considered single- and multi-pollutant models using linear regressions. Since Fast BKMR also allows for nonlinear effects, we modeled nonlinear NO$_2$, CO and PM$_{2.5}$ associations with natural cubic splines (3 degrees of freedom) in the single pollutant model, respectively. In the multi-pollutant models, we accounted for the nonlinear effects of each of the three pollutants simultaneously and included all possible combinations of pairwise interactions between NO$2$, CO, and PM$_{2.5}$. For Fast BKMR, we used 20 basis functions for the three pollutants. All models were adjusted for maternal race/ethnicity (white/black/Asian/Hispanic), age (modeled using a natural cubic spline with 5 df), education level (less than ninth grade, 9th–12th grade, high school, college and above), married status (married or unmarried), indicator of tobacco use during the pregnancy (yes or no),  indicator of alcohol use during the pregnancy (yes or no), indicator of previous birth (yes or no), indicator of gestational weeks, indicators for county of residence at delivery, Census track percent poverty levels (modeled using a natural cubic spline with 5 df) and conception date (modeled using natural cubic spline with 12 df). 

\subsection{Results}

The overall effects of all pollutants on birthweight differences, measured in grams, were estimated for two exposure contrasts: (1) between the $75^{th}$ and $25^{th}$ percentiles and (2) between the $95^{th}$ and $50^{th}$ percentiles of all exposures (Table~\ref{table:model compare}). In both single- and multi-pollutant models, the birthweight reduction from the exposure contrast of $95^{th}$ to $50^{th}$ percentiles were consistently smaller than that from the exposure contrast of $75^{th}$ to $25^{th}$ percentiles. However, unlike the linear models, Fast BKMR showed similar birthweight reductions for both exposure contrasts. Regardless of the exposure contrast, the single pollutant model consistently estimated smaller reduction in birthweight associated with air pollution. For the exposure contrast between $75^{th}$ and $25^{th}$ percentiles, multi-pollutant models with interaction terms had the largest effect sizes, with birthweight reductions reaching up to 25.06 grams ($95\%$ confidence interval (CI): 15.91, 34.21). For Fast BKMR, the reduction in birthweight was 18.18 ($95\%$ posterior interval (PI): 11.48, 24.87) grams, which is smaller than the effects size from multi-pollutant model but with a narrower $95\%$ PI. The difference in birthweight reduction between multi-pollutant models and Fast BKMR can be caused by the collinearity among exposures and different ways that exposure effects are aggregated. For the exposure contrast between $95^{th}$ and $50^{th}$ percentiles, the birthweight reduction from Fast BKMR was 16.24 grams ($95\%$ PI: 6.41, 26.08), while multi-pollutant models estimated a smaller reduction in birthweight.

\begin{table*}
\centering
\caption{Difference in average birthweight with 95\% confidence/posterior intervals (CI/PI) from single, multi-pollutant models and Fast BKMR. Effect sizes are evaluated with two different exposure contrasts with all exposures set at (1) the $75^{th}$ versus the $25^{th}$ percentile and (2) the $95^{th}$ versus the $50^{th}$ percentile.}
\label{table:model compare}
\tabcolsep=0pt
\begin{tabular*}{\textwidth}{@{\extracolsep{\fill}}llcc@{}}
    \hline
    \textbf{Model} & \textbf{Pollutant} & \textbf{$75^{th}$  vs  $25^{th} \,\ (95\% \,\ \text{CI/PI})$} & \textbf{$95^{th}$ vs $50^{th} \,\ (95\% \,\ \text{CI/PI})$} \\
    \hline
    \multirow{3}{*}{Single pollutant} & ns(NO$_2$, 3) & -16.54 (-23.93, -9.14) & -9.41 (-16.88, -1.93)\\
    & ns(CO, 3) & -12.05 (-18.63, -5.48) & -7.52 (-14.05, -0.99) \\
    & ns(PM$_{2.5}$, 3) & -5.97 (-10.26, -1.68) & -3.09 (-8.12, 1.94)\\
    \hline
    \multirow{5}{*}{Multi-pollutant} & ns(NO$_2$, 3) + ns(CO, 3)  & \multirow{2}{*}{-24.43 (-33.49, -15.36)} & \multirow{2}{*}{-9.99 (-19.13, -0.84)} \\
     & + ns(PM$_{2.5}$, 3) & & \\
     & ns(NO$_2$, 3) + ns(CO, 3) + & \multirow{3}{*}{-25.06 (-34.21, -15.91)} & \multirow{3}{*}{-13.03 (-23.06, -3.00)}\\
     & ns(PM$_{2.5}$, 3) + & & \\
    & three pairwise interactions &  &  \\
    \hline
    Fast BKMR &  $h(\text{NO}_2,\text{CO},
    \text{PM}_{2.5})$ & -18.18 (-24.87, -11.48) & -16.24 (-26.08, -6.41) \\
    \hline
    \end{tabular*}
\end{table*}

\begin{figure}
\centering
\includegraphics{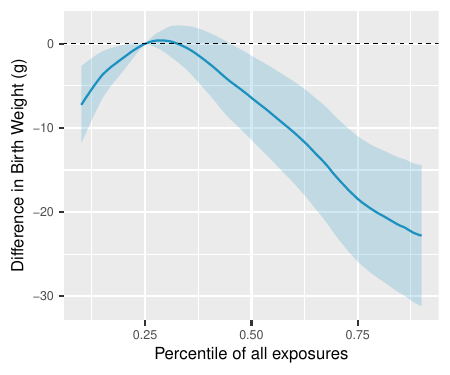}
\caption{The overall effects of all exposures on the birthweight compared to the 25$^{\text{th}}$ percentiles of exposures with 20 basis functions. Blue-shaded areas are the 95\% posterior intervals.}
\label{figure:overall effect}
\end{figure}

Figure~\ref{figure:overall effect} shows the overall effects of all three pollutants on birthweight differences, setting all three pollutants to their $25^{th}$ percentiles as the reference level. As the percentiles of the pollutants increased, birthweight initially increased slightly, followed by a large decrease. When three pollutants were at their $50^{th}$ percentiles, the reduction in average birthweight was around 7 grams. When all pollutants further increased to $95^{th}$ percentiles, the reduction exceeded 20 grams. The blue-shaded areas represent the 95\% posterior intervals. If the shaded region does not overlap with zero, it indicates a significant reduction in birthweight. We observed a significant reduction in birthweight when all exposures exceeded the $50^{th}$ percentile compared to when all exposures were at the $25^{th}$ percentile.

To explore the single pollutant effects, we constructed univariate exposure-response functions for each single pollutant while fixing the other two pollutants at $10^{th}, 50^{th}$ and $90^{th}$ percentiles. The possible range of each single pollutant was determined by varying the other two pollutants between the $(P - 5)^{th}$ and $(P + 5)^{th}$ percentiles, where $P$ represents the fixed percentiles of the two pollutants. For example, when we fixed CO and NO$_2$ at their $10^{th}$ percentiles, we generated response function for the observed range of PM$_{2.5}$ concentrations where CO and NO$_2$ were between their $5^{th}$ and $15^{th}$ percentiles.

\begin{figure*}
\centering
\includegraphics{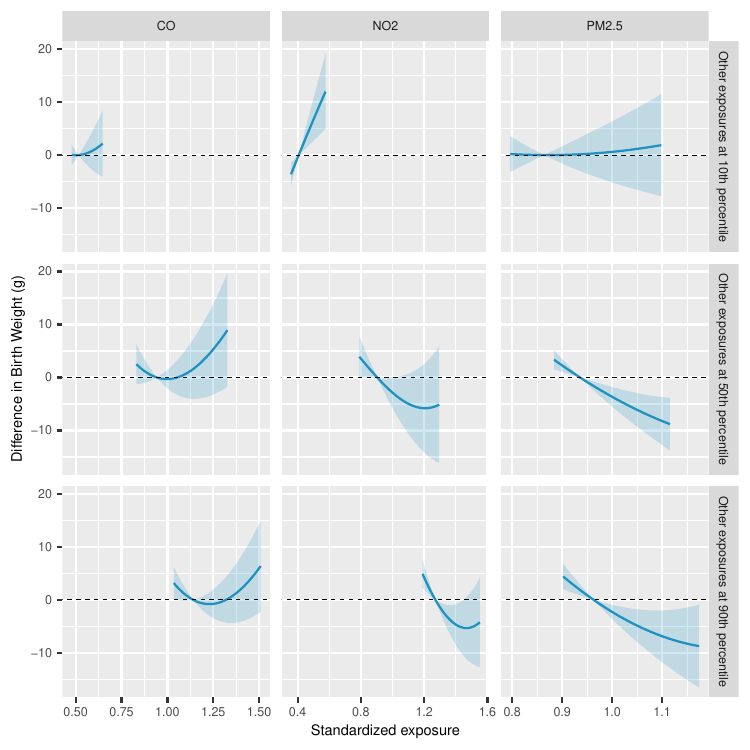}
\caption{The response plots for single pollutant when fixing other two pollutants at $10^{th}, 50^{th}$ and $90^{th}$ percentiles.}
\label{figure: single effect}
\end{figure*}

As shown in Figure~\ref{figure: single effect}, we found that increasing PM$_{2.5}$ was significantly associated with decreasing birthweight when CO and NO$_2$ were fixed at their $50^{th}$ and $90^{th}$ percentiles. For NO$_2$, we also observed birthweight reduction with increasing NO$_2$ levels when the other two pollutants were fixed at $50^{th}$ and $90^{th}$ percentiles, although this reduction had larger posterior intervals. In contrast, a reverse association was observed for CO, where birthweight increased with higher CO levels, regardless of the percentiles at which the other two pollutants were fixed.

\begin{figure*}
\centering
\includegraphics{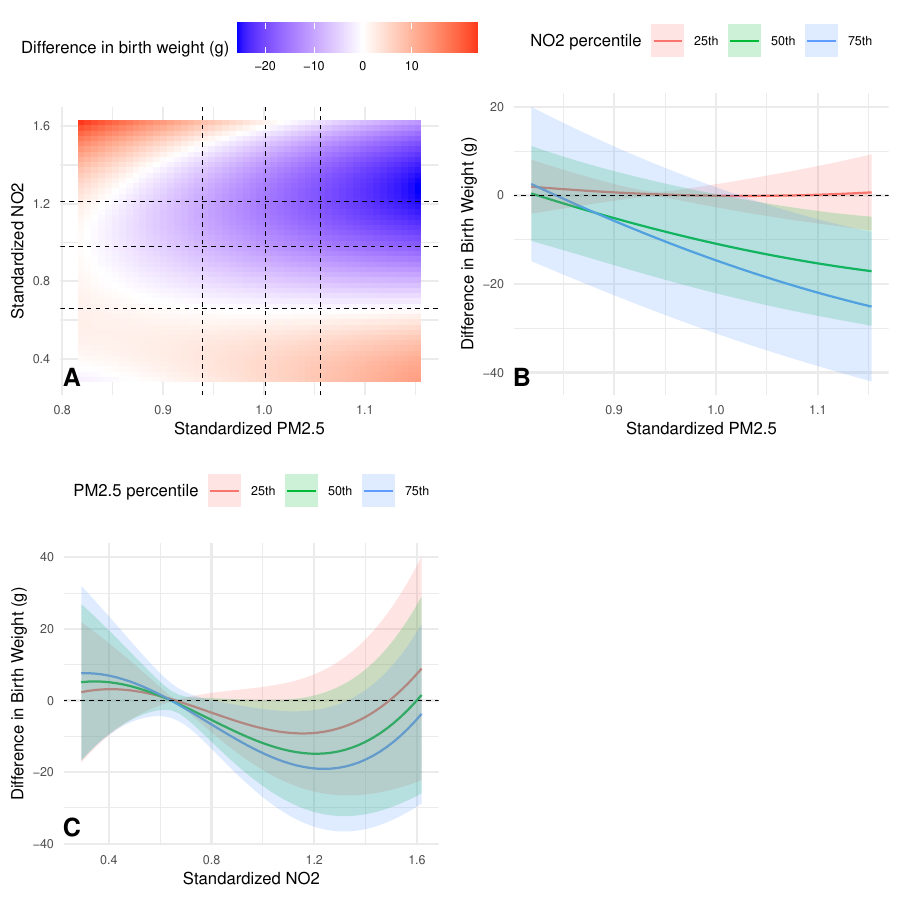}
\caption{\textbf{A.} Bi-pollutant exposure-response surface for standardized PM$_{2.5}$ and NO$_2$ when fixing the CO at $50^{th}$ percentiles. Vertical dashed lines are $25^{th}, 50^{th} ~\text{and}~ 75^{th}$ percentiles of standardized PM$_{2.5}$ (from left to right). Horizontal dashed lines are $25^{th}, 50^{th} ~\text{and}~ 75^{th}$ percentiles of standardized NO$_2$ (from bottom to top). \textbf{B.} Bi-pollutant exposure-response functions for standardized PM$_{2.5}$ at $25^{th}, 50^{th} ~\text{and}~ 75^{th}$ percentiles of the NO$_2$, with CO fixed at $50^{th}$ percentiles. \textbf{C.} Bi-pollutant exposure-response functions of standardized NO$_2$ with $25^{th}, 50^{th} ~\text{and}~ 75^{th}$ percentiles of the PM$_{2.5}$ with CO fixed at $50^{th}$ percentiles.}
\label{figure:heat map}
\end{figure*}

Finally, we explored the joint effects between two pollutants, fixing the third pollutant at its $50^{th}$ percentiles. The difference in birthweight was compared to the birthweight level when both two pollutants were at their $25^{th}$ percentiles. As shown in Figure~\ref{figure:heat map} A, the bi-pollutant exposure-response surface indicated that birthweight decreased as NO$_2$ and PM$_{2.5}$ concentrations increased together. The reduction in birthweight exceeded 20 grams when standardized PM$_{2.5}$ was over 1.1 and NO$_2$ around 1.2. When PM$_{2.5}$ was fixed at specific concentration, birthweight reduced with increasing NO$_2$, except when PM$_{2.5}$ concentration is also low. When standardized NO$_2$ was above 0.8, increasing PM$_{2.5}$ was consistently associated with a reduced birthweight. To make the bi-pollutant exposure-response surface more accessible, we present bi-pollutant exposure-response functions for pairwise exposures, where the second exposure is fixed at the $25^{th}, 50^{th}, ~\text{and} 75^{th}$ percentiles, in Figures~\ref{figure:heat map} B and C. The estimated difference in birthweight in Figures~\ref{figure:heat map} B and C is the same as the values represented by dashed lines in Figures~\ref{figure:heat map} A, but we also provide the 95\% posterior intervals for inference. As shown in Figure~\ref{figure:heat map} B, increasing PM$_{2.5}$ is associated with a greater reduction in birthweight at higher percentiles of NO$_2$. This reduction is significant when standardized PM$_{2.5}$ exceeds 1.05, with NO$_2$ fixed at the 50th and 75th percentiles. In Figure~\ref{figure:heat map} C, we observe a similar pattern, where a greater reduction in birthweight is associated with higher percentiles of PM$_{2.5}$ and increasing NO$_2$ levels between 0.8 and 1.2. This reduction is significant when NO$_2$ is fixed at the 75th percentile. The bi-pollutant exposure-response surfaces for standardized CO and PM$_{2.5}$ and for standardized CO and NO$_2$ are shown in Web Figures 4 and 5, respectively.

\section{Discussion}

In this study, we used supervised RFF to approximate Gaussian process, addressing the issue of computation time for BKMR with large datasets. According to the simulation studies, our methods can reduce computation time considerably, depending on the number of basis functions. Our approach also outperforms the use of predictive process in speed and estimation. Additionally, when sample size is small where the BKMR is feasible, our approach may also improve the estimation when the kernel induces strong correlation in the exposure-response surface, which is commonly observed among air pollutants. In scenarios with incorrect kernel specification, our proposed method is still able to capture the exposure-response relationships. In scenarios with parametric $h(\bx_i)$, our method performed comparable with BKMR with reduced computation time. Finally, our approach can also estimate marginal, single- and two-exposure exposure-response functions much faster than BKMR because matrix inversion is not needed in prediction.

Our findings on the effects of individual pollutants on birthweight, while holding the other two exposures constant, align with previous literature, showing that both NO$_2$ and PM$_{2.5}$ are associated with a reduction in birthweight (\cite{10.1001/jamanetworkopen.2020.8243,LI2019629,strickland2019associations}). According to the univariate exposure-response plots, we found positive association between CO and birthweight but this association is not statistically significant. To our knowledge, this is the first large-scale analysis of joint air pollutant effects on birthweight from administrative data using BKMR, offering new insights into the nonlinear and joint effects of air pollutants. In addition to using 20 basis functions, we also examined 5 and 50 basis functions. The overall and single exposure-response plots using 50 basis functions showed associations very similar to those observed with 20 basis functions. The exposure-response plots with 5 basis functions appeared closer to a straight line, indicating few basis functions may not be enough. The selection of the number of basis functions depends on various factors, including the number of exposures, sample size, and correlations between exposures. It is advisable to begin with a small number of basis functions and evaluate different choices to determine the optimal number.

Another advantage of using a low rank method is the ability to handle highly correlated random effects. In Gaussian process regression, the kernel matrix can become numerical singular (\cite{kanagawa2018gaussian}). This issue is common in environmental health studies on air pollution, where people living in the same areas, such as Census tracts or ZIP codes, often share the same exposure information, making them perfectly or highly correlated. By avoiding the inversion of the kernel matrix in our approach, we eliminate this issue.

One extension of our proposed method is to consider variable selection and dimension reduction. Specifically, stochastic variable selection can be conducted to exclude exposures from contributing to the kernel function. In the similar linear mixed-effect model framework, this is accomplished by placing a spike-and-slab prior on kernel parameter $\theta_m$ around zero (\cite{10.1111/j.0006-341X.2003.00089.x,10.1214/009053604000001147}). The current implementation of BKMR allows for variable selection, offering an approach for dimension reduction (\cite{10.1093/biostatistics/kxu058}). BKMR introduced variable selection to deal with  highly correlated exposures while our method  handles highly correlated exposures by learning the optimal frequencies for a small number of basis functions. Moreover, in population-based health studies examining the effects of multiple air pollutants or the effects of chemicals with extensive biomonitoring, the number of exposures is typically much smaller than the sample size. Hence, dimensionality reduction is not always needed to characterize the exposure-response function. 

There are other further methodological investigations for the Fast BKMR. First, compared with non-Bayesian methods, such as WQS and q-gcomp, the Fast BKMR is still time consuming. Since BKMR is preferred in scenarios considering non-linear effects and interactions, we may consider approaches to improve the computation of BKMR without MCMC. Second, our current method only works for continuous outcomes. One potential extension is to apply this method in a generalized linear model framework, where the outcome can follow binomial or Poisson distributions.

\section{Acknowledgments}
The first and third authors were supported by grants from the National Institute of Environmental Health Sciences (NIEHS) under Award Number R01ES028346. The second author was support by NIEHS under Award Number K01ES035082. Any opinions, findings, and conclusions or recommendations expressed in this material are those of the authors and do not necessarily reflect the views of the funders.


\section{Web supplement}
The web supplement contains an appendix with HMC update details and additional results from simulation studies and applications. \\
R code used to implement the proposed methods is available at \url{https://github.com/Danlu233/fastbkmr}.




\end{document}


\maketitle

\section{Hamiltonian Monte Carlo (HMC) updates}

For regression coefficients $\Theta = (\gamma_1, \ldots, \gamma_P, a_1, \ldots, a_J, b_i, \ldots, b_J)^T$ updates, let $\bB$ be the $n\times (P+2J)$ design matrix with each row having elements
\[ [Z_{i1}, \ldots, Z_{iP}, \cos (\bomega^T_1\bx_i), \ldots, \cos (\bomega^T_J\bx_i), \sin (\bomega^T_1\bx_i), \ldots, \sin (\bomega^T_J\bx_i)] \]  Then $\mathcal{L}(\Theta)$ and $\nabla\mathcal{L}(\Theta)$ used in HMC is \\
\[ \mathcal{L}(\Theta) = -\frac{1}{2\sigma^2} (\bY - \bB\Theta)^T(\bY - \bB\Theta)-\frac{1}{2}\Theta^T\bS^{-1}\Theta \] and \\ 
\[ \nabla_{\Theta} \mathcal{L}(\Theta) = \frac{1}{\sigma^2} \bB^T (\bY - \bB\Theta)-\bS^{-1} \Theta \] \\
where $\bS = diag (\sigma^2_{\bgamma}, \ldots, \sigma^2_{\bgamma}, \tau^2/J\, \ldots, \tau^2/J)$. 

For $\bomega_j$ updates, let $\bB_{cos}$ be the design matrix with rows $[\cos (\bomega^T_1\bx_i), \ldots, \cos (\bomega^T_J\bx_i)]$, and similarly define $\bB_{sin}$. Then 
\[ \mathcal{L}(\Omega) = -\frac{1}{2\sigma^2} ||\bY - \bB_{cos}\ba - \bB_{sin}\bb - \bZ^T\bgamma ||^2-\frac{1}{2}\sum_{j=1}^J\bomega_j^T\bSigma^{-1}\bomega_j\]
\[ \nabla_{\Omega}\mathcal{L}(\Omega)_{J\times M} = -\frac{1}{\sigma^2} \bD^T (\bX \odot [\bR \,|\, \ldots \,|\, \bR])- \Omega^*\]
where $\odot$ is the Hadamard (element-wise) product, and 
\[ \bD = \bB_{sin} \odot [\ba \,|\, \ba \,|\, \ldots \,|\, \ba]^T 
- \bB_{cos} \odot [\bb \,|\, \bb \,|\, \ldots \,|\, \bb]^T\]
\[ \bR = \bY - \bB_{cos}\ba - \bB_{sin}\bb - \bZ\bgamma\]
\[\Omega^* = [\bSigma^{-1}\bomega_1 \,|\, \bSigma^{-1}\bomega_2 \,|\, \ldots \,|\, \bSigma^{-1}\bomega_J ]^T   \]

\section{Tables and Figures}

\begin{table}
    \caption{Simulation settings for exposures and confounders.}
    \label{SI:set}
    \centering
    \begin{tabular}{c|c}
    \hline
       Simulated Variable  & Distribution \\
       \hline
       Exposure 1  & Normal(0, $0.9^2$) \\
       Exposure 2  & Normal(0, $2.4^2$) \\
       Exposure 3  & Normal(0, $1.2^2$) \\
       Exposure 4  & Normal(0, $2.6^2$) \\
       Exposure 5  & Normal(0, $2.8^2$) \\
       Exposure 6  & Normal(0, $0.1^2$) \\
       Exposure 7  & Normal(0, $1.6^2$) \\
       Exposure 8  & Normal(0, $2.7^2$) \\
       Exposure 9  & Normal(0, $1.7^2$) \\
       Exposure 10  & Normal(0, $1.4^2$) \\
       Confounder 1 & Normal(3, $6^2$) \\
       Confounder 2 & Binominal(1, 0.7) \\
       Confounder 3 & Normal(2, $0.5^2$) \\
       Confounder 4 & Normal(1, $5^2$) \\
       Confounder 5 & Binominal(1, 0.3) \\
       \hline
    \end{tabular}
\end{table}

\begin{table*}
\caption{Computation time in minutes (percentages of reduction$^\text{1}$) for BKMR, BKMR with Predictive Process (PP) and Fast BKMR at sample sizes of 200 and 500.}
\label{SI:compute_time}
\tabcolsep=0pt
\begin{tabular*}{\textwidth}{@{\extracolsep{\fill}}lllll@{}}
\hline
Sample Size & Type & $M^\text{2} = 2$ & $M^\text{2} = 5$ & $M^\text{2} = 10$ \\
\hline
\multirow{6}{5em}{200} 
& BKMR & 8.69 (0\%)  &  7.61 (0\%)  &7.69 (0\%)   \\
& BKMR (PP) & 1.18 (86\%)  &  3.63 (52\%)  &9.4 (-22\%)   \\
& Fast BKMR (J = 5)   & 0.49 (94\%)  &  0.87 (89\%)  &0.97 (87\%)  \\
& Fast BKMR (J = 100) & 2.24 (74\%)  &  3.36 (56\%)  &3.24 (58\%)  \\
& Fast BKMR (J = 1000) & 20.43 (-135\%)  &  28.63 (-276\%)  &27.18 (-254\%)  \\
\hline
\multirow{6}{5em}{500} 
& BKMR  & 117.43 (0\%)  &80.3 (0\%)  &117.67 (0\%)  \\
& BKMR (PP) & 5.14 (96\%)  &12.36 (85\%)  &40.92 (65\%) \\
& Fast BKMR (J = 5)   & 0.68 (99\%)  &0.97 (99\%)  &1.46 (99\%)  \\
& Fast BKMR (J = 100)  & 5.64 (95\%)  &6.82 (92\%)  &8.21 (93\%)  \\
& Fast BKMR (J = 1000) & 57.51 (51\%)  &66.33 (17\%)  &76.69 (35\%)  \\
\hline
\end{tabular*}

\begin{tablenotes}
\item $^{1}$ Percentage of reduction: $1 - \frac{\text{computation time of current model}}{\text{computation time of BKMR}}$
\item $^{2}$ $M$: the number of exposures. 

\end{tablenotes}

\end{table*}

\begin{table}
    \centering
    \caption{Population Characteristics in Altanta Birthweight Study.}
    \label{SI:chara}
    \begin{tabular}{lc}
    \hline
    \textbf{Characteristics} & \textbf{Overall (N = 273,711)} \\
    \hline
    \textbf{Maternal race/ethnicity} & \\
    \hspace{5mm} White & 121,056 (44.2\%) \\
    \hspace{5mm} Black & 84,193 (30.8\%) \\
    \hspace{5mm} Asian & 13,114 (4.8\%) \\
    \hspace{5mm} Hispanic & 53,831 (19.7\%) \\
    \hspace{5mm} Other & 1,517 (0.6\%) \\
    \textbf{Maternal age} & \\
    \hspace{5mm} 16 - 25 years & 89,401 (32.7\%) \\
    \hspace{5mm} 25 - 31 years & 104,352 (38.1\%) \\
    \hspace{5mm} 31 - 43 years & 79,958 (29.2\%) \\
    \textbf{Maternal education} & \\
    \hspace{5mm} Less than 9th grade & 22,656 (8.3\%) \\
    \hspace{5mm} 9th-12th grade & 39,236 (14.3\%) \\
    \hspace{5mm} High School & 76,481 (27.9\%)\\
    \hspace{5mm} College and above & 135,338 (49.4\%) \\
    \textbf{Maternal tobacco use} &  12,644 (4.6\%)\\
    \textbf{Maternal alcohol use} & 1,723 (0.6\%) \\
    \textbf{Maternal married} & 174,009 (63.6\%) \\
    \textbf{Previous birth} &  164,202 (60.0\%)\\
    \textbf{Gestational weeks} & \\
    \hspace{5mm} Mean (SD) & 38.7 (1.8) \\
    \hspace{5mm} [Min, Max] & [28.0, 44.0] \\
    \textbf{Census block group percent poverty levels} &\\
    \hspace{5mm} Mean (SD) & 0.10 (0.10) \\
    \hspace{5mm} [Min, Max] & [0, 0.77] \\
    \hline
    \end{tabular} \\
    \vspace{1mm}
\end{table}

\begin{table}
    \centering
    \caption{Statistics Summary of Pollutants in Atlanta Birthweight Study (original scale).}
    \label{SI:exposure}
    \begin{tabular}{cccc}
    \hline
    Pollutant & Mean (SD) & [Min, Max] & IQR \\
    \hline
        NO$_2$ (ppb) & 22.03 (7.86) & [3.88, 40.47] & 12.96\\
        CO (ppm) & 0.70 (0.23) & [0.23, 1.41] & 0.34\\
        PM$_{2.5}$ ($\mu \text{g/m}^3$) & 15.90 (1.24) & [11.19, 20.66] & 1.86\\
        \hline
    \end{tabular}
    
\end{table}

\begin{figure}[h]
\centering
\includegraphics{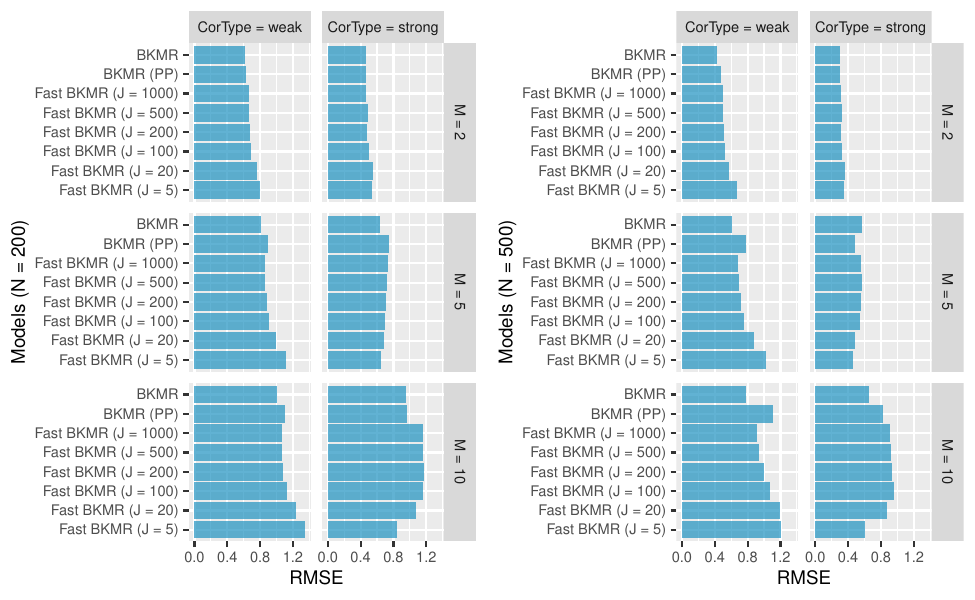}
\caption{The RMSE between true and estimated joint effects with different correlation type (CorType), number of exposures ($M$) and sample size ($N$).}
\label{SI:HRMSE}
\end{figure}

\begin{figure}[h]
\centering
\includegraphics{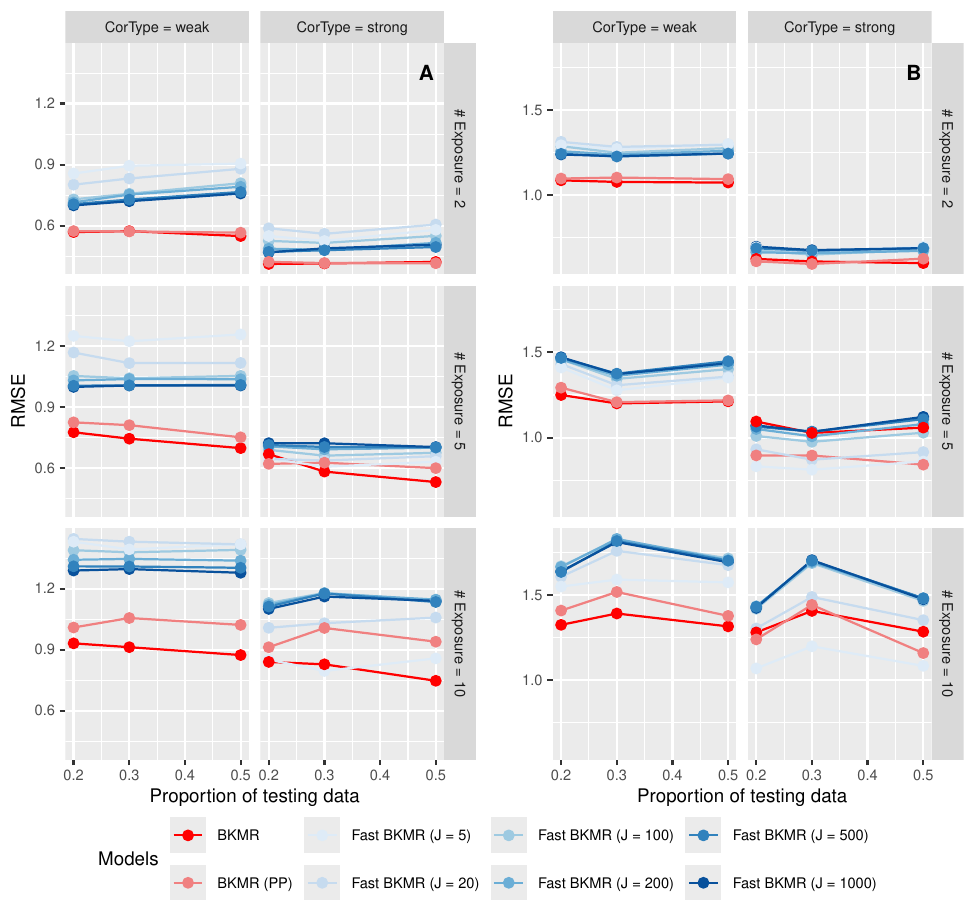}
\caption{The RMSE from out-of-sample test on 200 observations with different correlation type (CorType), number of exposures and kernels. $\mathbf{A}$: Correct specified kernel. $\mathbf{B}$: Incorrect specified kernel with $\exp (- \sum_{m=1} ^M \theta_m \sqrt{|x_{im} - x_{jm}|})$.}
\label{SI:oost}
\end{figure}

\begin{figure}[h]
\centering
\includegraphics{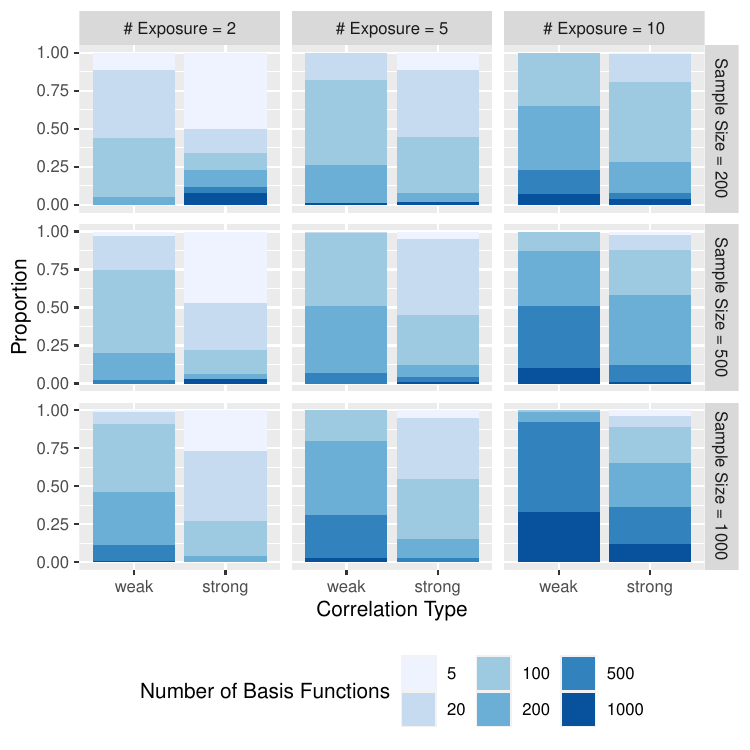}
\caption{The proportion of selected number of basis functions based on WAIC.}
\label{SI:WAIC}
\end{figure}

\begin{figure}[h]
\centering
\includegraphics{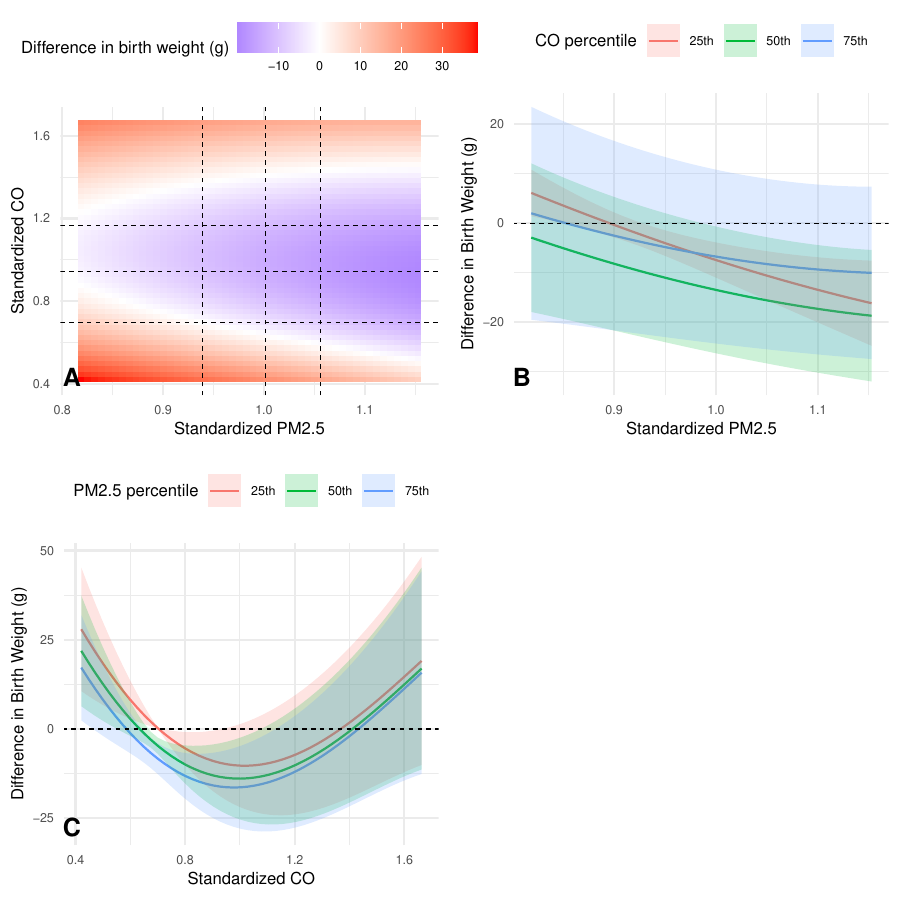}
\caption{\textbf{A.} Bi-pollutant exposure-response surface for standardized CO and PM$_{2.5}$ when fixing the NO$_2$ at $50^{th}$ percentiles. Vertical dashed lines are $25^{th}, 50^{th} ~\text{and}~ 75^{th}$ percentiles of standardized PM$_{2.5}$ (from left to right). Horizontal dashed lines are $25^{th}, 50^{th} ~\text{and}~ 75^{th}$ percentiles of standardized CO (from bottom to top). \textbf{B.} Bi-pollutant exposure-response functions for standardized PM$_{2.5}$ at $25^{th}, 50^{th} ~\text{and}~ 75^{th}$ percentiles of the CO, with NO$_2$ fixed at $50^{th}$ percentiles. \textbf{C.} Bi-pollutant exposure-response functions of standardized CO with $25^{th}, 50^{th} ~\text{and}~ 75^{th}$ percentiles of the PM$_{2.5}$ with NO$_2$ fixed at $50^{th}$ percentiles.}
\label{SI:heat_map_fix_NO2}
\end{figure}

\begin{figure}[h]
\centering
\includegraphics{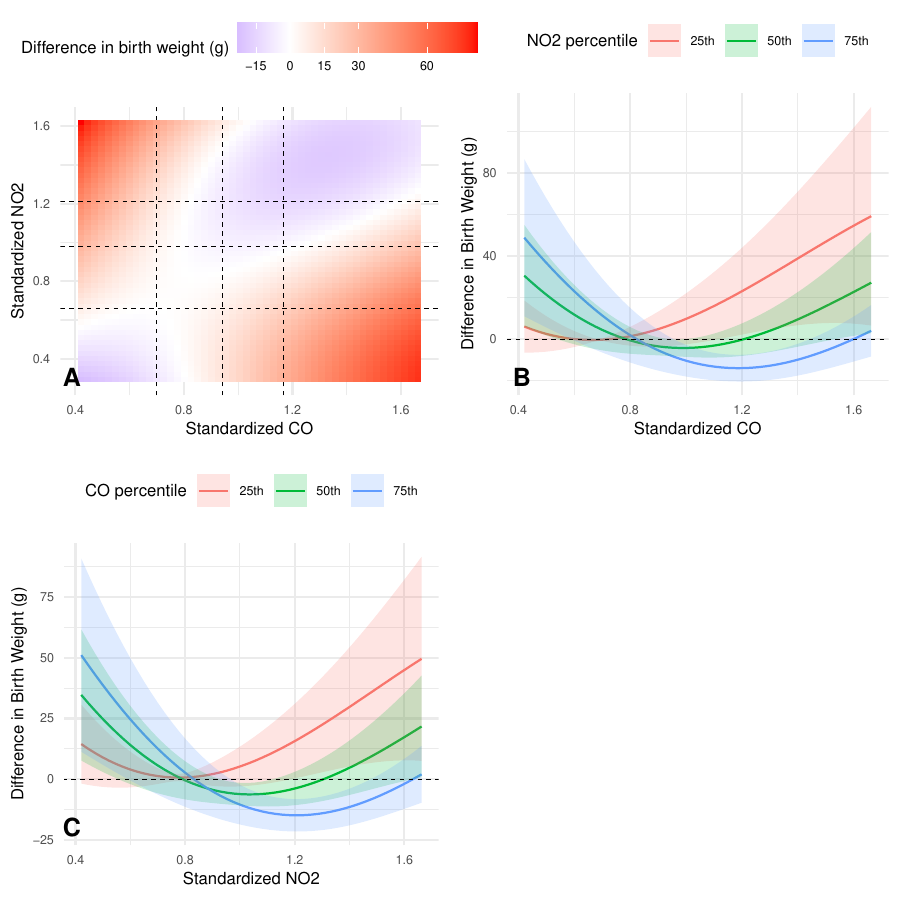}
\caption{\textbf{A.} Bi-pollutant exposure-response surface for standardized CO and NO$_2$ when fixing the PM$_{2.5}$ at $50^{th}$ percentiles. Vertical dashed lines are $25^{th}, 50^{th} ~\text{and}~ 75^{th}$ percentiles of standardized CO (from left to right). Horizontal dashed lines are $25^{th}, 50^{th} ~\text{and}~ 75^{th}$ percentiles of standardized NO$_2$ (from bottom to top). \textbf{B.} Bi-pollutant exposure-response functions for standardized CO at $25^{th}, 50^{th} ~\text{and}~ 75^{th}$ percentiles of the NO$_2$, with PM$_{2.5}$ fixed at $50^{th}$ percentiles. \textbf{C.} Bi-pollutant exposure-response functions of standardized NO$_2$ with $25^{th}, 50^{th} ~\text{and}~ 75^{th}$ percentiles of the CO with PM$_{2.5}$ fixed at $50^{th}$ percentiles.}
\label{SI:heat_map_fix_PM25}
\end{figure}

%% file: mydef.tex
\def\bSig\mathbf{\Sigma}

\newcommand{\bgamma}{ \mbox{\boldmath $\gamma$}}
\newcommand{\bomega}{ \mbox{\boldmath $\omega$}}

\newcommand{\bx}{ \mbox{\boldmath $x$}}

\newcommand{\bh}{ \mbox{\bf h}}

\newcommand{\bz}{ \mbox{\boldmath $z$}}

\newcommand{\bK}{ \mbox{\bf K}}

\newcommand{\bZero}{ \mbox{\boldmath $0$}}

\newcommand{\vtwo}{\vspace{.2cm}}